\documentclass[aps,prd,twocolumn,superscriptaddress,amsfont,graphicx,nofootinbib,preprintnumbers]{revtex4}%
\usepackage{color,graphicx,epsfig}
\usepackage{ifpdf}
\usepackage{amsmath}
\usepackage{bm}
\usepackage{color}
\usepackage[english]{babel}
\usepackage{graphicx}%
\usepackage{amsfonts}%
\usepackage{amssymb}
\usepackage{braket}
\usepackage{hyperref}

\definecolor{nicered}{rgb}{0.7,0.1,0.1}
\definecolor{nicegreen}{rgb}{0.1,0.5,0.1}
\hypersetup{colorlinks,citecolor= nicegreen,linkcolor= nicered}

\newcommand{\cB}{\mathcal B}
\newcommand{\cA}{\mathcal A}

\newcommand{\be}{\begin{equation}}
\newcommand{\ee}{\end{equation}}
\newcommand{\bea}{\begin{eqnarray}}
\newcommand{\eea}{\end{eqnarray}}

\definecolor{Red}{rgb}{1.,0.,0.}

\newcommand{\SM}{{\rm SM}}

\newcommand{\gsim}{\stackrel{>}{_\sim}}

\begin{document}

\preprint{\tiny CERN-PH-TH/2012-128}

\title{Shedding light on CP violation in the charm system via $D\to V \gamma$ decays}

\author{Gino Isidori}
\affiliation{CERN, Theory Division, CH1211 Geneva 23, Switzerland}
\affiliation{INFN, Laboratori Nazionali di Frascati, Via E. Fermi 40, 00044 Frascati, Italy}

\author{Jernej F.\ Kamenik} 
%\email[Electronic address:]{jernej.kamenik@ijs.si} 
\affiliation{J. Stefan Institute, Jamova 39, P. O. Box 3000, 1001 Ljubljana, Slovenia}
\affiliation{Department of Physics, University of Ljubljana, Jadranska 19, 1000 Ljubljana, Slovenia}

\date{\today}
\begin{abstract}
Recent evidence for direct CP violation in non-leptonic charm decays cannot be easily accommodated within the Standard Model (SM). On the other hand, it fits well in new physics models generating CP violating $\Delta C =1$ chromomagnetic dipole operators. We show that in these frameworks sizable direct CP asymmetries in radiative $D\to P^+ P^- \gamma$ decays ($P=\pi,K$), with $M_{PP}$ close to the $\rho$ or the 
$\phi$ peak, can be expected. 
% Enhanced matrix elements of the electromagnetic dipole operators can partly compensate the long distance dominance in these decays and 
% the resulting CP asymmetries can reach the order of several percent, well above what expected within the SM. 
% We briefly comment on related CP violating observables accessible via time dependent $D(\bar D)\to P^+ P^- \gamma$ studies and 
% angular decay product distributions in rare semileptonic $D$ decays.
Enhanced matrix elements of the electromagnetic dipole operators can partly compensate the long distance dominance in these decays, 
leading to CP asymmetries of the order of several percent. If observed at this level, these would provide a clean signal of physics beyond 
the SM and of new dynamics associated to dipole operators. 
We briefly comment on related CP violating observables accessible via time dependent $D(\bar D)\to P^+ P^- \gamma$ studies and angular decay product distributions in rare semileptonic $D$ decays.
\end{abstract}

\maketitle

\section{Introduction}

A significant evidence for direct CP violation in $D\to P^+P^-$ decays 
($P=\pi,K$) has recently been reported by the LHCb~\cite{Aaij:2011in} and by 
the CDF~\cite{Aaltonen:2011se} collaborations. Both experiments find a non-vanishing 
value for $\Delta a_{CP} \equiv a_{K^+ K^-} - a_{\pi^+ \pi^-}$, where
\begin{equation}
\label{eq:acpdef}
a_f \equiv \frac{\Gamma(D^0\to f)-\Gamma(\bar D^0\to f)}
  {\Gamma(D^0\to f)+\Gamma(\bar D^0\to f)}\,.
\end{equation}
Combining these recent results with older measurements 
leads to  the following world average~\cite{Aaltonen:2011se} 
\begin{equation}
\Delta a^{\rm exp}_{CP} = -(0.67\pm 0.16)\%\,,
\label{eq:acpExp}
\end{equation}
that differs from zero by about $4\sigma$. 

The theoretical interpretation of this result is puzzling. The value in Eq.~(\ref{eq:acpExp}) exceeds by a factor 5--10 what 
is naturally expected in the Standard Model (SM) (see e.g.~Ref.~\cite{Grossman:2006jg}, and the more recent 
analyses in Ref.~\cite{Isidori:2011qw,Cheng:2012wr}). However, we cannot exclude that such result
has a SM explanation due to the non-perturbative enhancement of penguin-type
hadronic matrix elements~\cite{Golden:1989qx, Brod:2011re,Pirtskhalava:2011va}.  
On the other hand, 
this value can naturally be accommodated  in well-motivated extensions of the SM. In particular, it fits well in models 
generating at short distances a sizable CP violating phase for the effective $\Delta C=1$ chromomagnetic 
operators~\cite{Grossman:2006jg,Isidori:2011qw,Giudice:2012qq,Altmannshofer:2012ur}.

Given this situation, it is important to identify possible future experimental tests able to distinguish 
standard vs.~non-standard explanations of $\Delta a_{CP}$. An interesting strategy that makes use 
of CP asymmetries in various hadronic $D$ decays (necessarily including neutral mesons) has recently been 
proposed in Ref.~\cite{Grossman:2012eb}. However, this strategy is effective in  isolating possible non-standard
contributions to  $\Delta a_{CP}$ only if they are generated by effective operators with a $\Delta I=3/2$ isospin structure.
This is not  the case for the well-motivated scenario with a new CP violating phase  in the $\Delta C=1$ chromomagnetic  operator. As we point out here, in the latter case an efficient strategy 
is obtained by measuring CP asymmetries in radiative $D$ decays. 

\section{Short-distance effective Hamiltonian}

The first key ingredient of our strategy is the strong link between the $\Delta C=1$ chromomagnetic operator, 
\bea
\mathcal Q_{8} &=&  \frac{m_c}{4\pi^2}\, \bar u_L \sigma_{\mu\nu} 
    T^a g_s G_a^{\mu\nu} c_R \,, 
   \label{eq:Q8def}
\eea
and the $\Delta C=1$ electromagnetic-dipole operator, 
\bea
\mathcal Q_{7} &=&  \frac{m_c}{4\pi^2}\, \bar u_L \sigma_{\mu\nu} 
  Q_u  e F^{\mu\nu} c_R \,.
   \label{eq:Q7def}
\eea
In most explicit new-physics models the short-distance Wilson coefficients  of these two operators ($C_{7,8}$) are expected to be 
similar. Moreover, even assuming that only a non-vanishing $C_{8}$ is generated at some high scale, 
the mixing of the two operators under the QCD renormalization group (RG) implies $C_{7,8}$ of comparable size at the charm scale.
The same is true for the pair of operators with opposite chirality $\mathcal Q^\prime_{7,8}$, obtained from  $\mathcal Q_{7,8}$ 
with the replacement $L \leftrightarrow R$. 

To quantify the size of these coefficients, we normalize the effective Hamiltonian describing the  $\Delta C=1$ 
new-physics contributions as 
\be
\mathcal H^{\rm eff-\rm NP}_{|\Delta c| = 1} = \frac{G_F}{\sqrt 2} 
  \sum_{i} C_i  \mathcal Q_i + {\rm h.c.}\,,
\label{eq:HNP}
\ee
The complete list of potentially relevant operators can be found in Ref.~\cite{Isidori:2011qw}; however, for the purpose of our 
analysis we can restrict our attention only to $\mathcal Q_{7,8}$ and $\mathcal Q^\prime_{7,8}$. Assuming the initial conditions of these operators 
are generated at some scale $M > m_t$,  taking into account the RG evolution of the operators at the leading log 
level (assuming only SM degrees of freedom below the scale $M$), leads to~\cite{Buras:1999da}
\begin{eqnarray}
C^{(\prime)}_7(m_c) &=& \tilde \eta  \left[ \eta C^{(\prime)}_7(M) + 8\,
  (\eta -1)\, C^{(\prime)}_8(M) \right], \\ 
C^{(\prime)}_8(m_c) &=& \tilde \eta\, C^{(\prime)}_8(M),
\end{eqnarray}
where
\begin{equation}\label{eta}
\eta=\left[\frac{\alpha_s(M)}{\alpha_s(m_t)}\right]^\frac{2}{21}
        \left[\frac{\alpha_s(m_t)}{\alpha_s(m_b)}\right]^\frac{2}{23}
        \left[\frac{\alpha_s(m_b)}{\alpha_s(m_c)}\right]^\frac{2}{25}~,
\end{equation}
and
\begin{equation}
\tilde \eta = \left[ \frac{\alpha_s(M)}{\alpha_s(m_t)}\right]^{\frac{14}{21}} \left[ \frac{\alpha_s(m_t)}{\alpha_s(m_b)}\right]^{\frac{14}{23}} \left[ \frac{\alpha_s(m_b)}{\alpha_s(m_c)}\right]^{\frac{14}{25}}\,.
\end{equation}

Following the analysis in Ref.~\cite{Giudice:2012qq}, the new-physics contribution to 
$\Delta a_{CP}$ induced by $\mathcal Q_{8}$ can be written as 
\be
| \Delta a^{\rm NP}_{CP} | \approx - 1.8   |\mathrm{Im}[C^{\rm NP}_8(m_c)] |~,
\label{eq:acpNP}
\ee
where the numerical value assumes maximal strong phases and is affected by $\mathcal O(1)$ 
uncertainties due the theoretical error on
 $\langle PP |   \mathcal Q_8 | D\rangle$.
Assuming this contribution saturates the experimental value of $\Delta a_{CP}$
leads to
$ |\mathrm{Im}[C^{\rm NP}_8(m_c)] | \approx  0.4 \times 10^{-2}$. 
If we further assume that the initial scale $M$ is around 1 TeV, and that at this scale
$|C^{\rm NP}_7(M)| \ll |C^{\rm NP}_8(M)|$, the RG evolution implies
\be
|\mathrm{Im}[C^{\rm NP}_7(m_c)] | \approx  |\mathrm{Im}[C^{\rm NP}_8(m_c)] |
\approx  0.4 \times 10^{-2}~. 
\label{eq:C7expect}
\ee
This is for instance what happens in supersymmetry, where the gluino-mediated 
amplitude proportional to $(\delta^D_{LR})_{12}$ leads to the 
initial condition
\be
 C^{\rm SUSY}_7(m_{\rm SUSY}) =  (4/15) C^{\rm SUSY}_8(m_{\rm SUSY})~.
 \ee
 Taking into account the $\mathcal O(1)$ 
uncertainties in the determination of $ |\mathrm{Im}[C^{\rm NP}_8(m_c)] |$, and the 
additional uncertainties in the initial conditions of $C^{\rm NP}_7(M)$,   
we consider the following range for Im$(C^{\rm NP}_7)$ at the charm scale
\be
  |\mathrm{Im}[C^{\rm NP}_7(m_c)] | = (0.2-0.8) \times 10^{-2}~.
\ee
The same range holds for Im$(C^\prime_7)$, if the leading contribution to 
$\Delta a_{CP}$ is generated by $\mathcal Q_8^\prime$ rather than $\mathcal Q_8$ . 

At low energies $C_7$ receives contributions also from the mixing with 
the SM four-fermion operators.  However, to a good accuracy 
these contributions are CP conserving. The leading effect is the 
two-loop mixing between $C_7$ and $C^{s,d}_{1,2}$~\cite{Greub}.
According to the analysis in Ref.~\cite{Greub},
integrating out also light quark loops one obtains 
\be
|C^{\rm SM-eff}_7(m_c) | =  (0.5 \pm 0.1) \times 10^{-2}~,
\label{eq:C7SM}
\ee
with an $\mathcal O(1)$ strong phase and a negligible CP-violating phase
(more than two orders of magnitude smaller).

If the contributions in Eqs.~(\ref{eq:C7expect}) and (\ref{eq:C7SM}) where
the dominant contributions to radiative $D$ decays, we could expect 
$\mathcal O(1)$ direct CP asymmetries in these modes. As we discuss below, 
this is not the case due  to genuine long-distance 
contributions that dominate the decay rates.

\section{Short- vs.~long-distance contributions in $D\to V\gamma$}

The second important ingredient of our analysis is the observation 
that in the Cabibbo-suppressed $D\to V\gamma$ decays, 
where $V$ is a light vector meson with $u\bar u$ valence quarks 
($V=\rho^0,\omega$),  $Q_7$ and $Q_7^\prime$ have a sizable hadronic matrix element.
More explicitly, the short-distance contribution induced by $Q^{(\prime)}_7$, relative to the total (long-distance) amplitude,  
is substantially larger with respect to the corresponding relative weight of $Q^{(\prime)}_8$ in $D\to P^+P^-$ decays.

The decay amplitudes for $D\to V\gamma$ decays can be decomposed as follows 
\bea
&& \cA[ D(p) \to V(\tilde p,\tilde \epsilon) \gamma(q,\epsilon) ] =  
 -i A^V_{\rm PC}~ \epsilon_{\mu\nu\alpha\beta} q^\mu\epsilon^{*\nu}p^\alpha \tilde \epsilon^\beta   \nonumber \\
&& \quad  +A^V_{\rm PV} ~ [ (\tilde \epsilon^* q )(\epsilon^* p)-(qp) (\tilde \epsilon^* \epsilon^* ) ]~, \qquad 
\eea
with the corresponding rates
\be
\Gamma(D\to V\gamma) = \frac{m_D^3}{32\pi} \left(1 - \frac{m_V^2}{m_D^2}\right)^3   \left[ |A_{PV}|^2 + |A_{PC}|^2 \right]~.
\ee
The short-distance contribution induced by $Q_7$ to the effective couplings $A^V_{\rm PV,PC}$ is
\be
(A^{V}_{\rm PC(PV)})^{\rm s.d.} = \frac{e Q_u G_F}{\sqrt{2}} \frac{ m_c  }{2\pi^2}  C_7(m_c) ~T^{V}_{1(2)},
\ee
where $T^V_{1(2)}$ are defined by 
\bea
&& \langle  V(\tilde p,\tilde \epsilon) | \bar u  q_\nu \sigma^{\mu\nu} (1+ \gamma_5) c | D(p) \rangle 
= -2 i \epsilon^{\mu\alpha \beta\sigma} \tilde\epsilon^{*\alpha} p^\beta \tilde p^\sigma   T^V_1 \nonumber\\
&&\qquad + T^V_2 \left[ (m_D^2 -m_V^2)\tilde\epsilon^{*\mu} -(\tilde \epsilon^* p) (p+\tilde p)^\mu \right]~,
\label{eq:MatQ7}
\eea
and $T^V_1=T^V_2 \equiv T_{(D)}^V$ via the identity $\gamma_5 \sigma^{\mu\nu} = \frac{i}{2} \varepsilon^{\mu\nu\alpha\beta} \sigma_{\alpha\beta}$.
A recent sum-rule estimate finds~\cite{sumrules} $T_{(D)}^\rho \approx T_{(D)}^\omega \approx 0.70(7)$.
We note in passing that at leading order in $\alpha_s$ and in the infinite charm quark mass limit, heavy quark symmetry predicts $T^V_{(D)} = V^{V}_{(D)}(0)/(1-m_V/m_D)$~\cite{LEET}, where ($q^2 \equiv (p-\tilde p)^2$)
\be
\langle V(\tilde p, \tilde \epsilon) | \bar u \gamma^\mu c | D(p) \rangle = 2i \frac{V^V_{(D)}(q^2)}{m_D+m_V} \varepsilon^{\mu\nu\alpha\beta} p_\nu \tilde p_\alpha \tilde \epsilon_\beta\,.
\ee
This matrix element enters semileptonic $D\to V$ decays and thus $V^V_{(D)}(0)$ can be accessed experimentally. Unfortunately for the interesting $D\to (\rho,\omega) \ell \nu$ transitions, no such analyses are available at present.
On the other hand, in the heavy charm quark limit, $T^V_{(D)}$ can be related to hadronic matrix elements entering radiative $B\to V$ transitions -- $T_{(B)}^V$.  Starting from the (quenched) Lattice QCD estimate $T^\rho_{(B)} = 0.20(4)$~\cite{Becirevic:2006nm}, and running it in both the perturbative matching scale (from $\mu_b=4.6$~GeV to $\mu_c = 1.4$~GeV) as well as  the heavy quark mass scaling (including leading power corrections)~\cite{Becirevic:2006nm,Damir}, we obtain $T_{(D)}^\rho \approx 0.7(2)$.
%together with the leading heavy quark effective theory (HQET) scaling of $T^V_{(H)} \sim m_H^{-3/2}$, we obtain $T_{(D)}^\rho \approx 1.0(2)$. Leading power corrections to the HQET scaling of $T^V_{(H)}$ have also been studied~\cite{Becirevic:2006nm,Damir} and lead to a smaller estimate of $T_{(D)}^\rho \approx 0.7(2)$. 
Using instead existing sum-rule estimates of $T^V_{(B)}$~\cite{sum-rule-B} typically leads to $\mathcal O(20\%)$ larger  values. Consequently we employ the value of $T^V_{(D)}$ with a conservative uncertainty estimate of
\be
T_{(D)}^\rho \approx T_{(D)}^\omega \approx 0.7\pm 0.2\,,
\ee
which leads to
\be
\left|(A^{\rho,\omega}_{\rm PC,PV})^{\rm s.d.} \right|
\approx  \frac{0.6(2) \times 10^{-9} }{ m_D}  \left| \frac{ C_7(m_c) }{ 0.4 \times 10^{-2} } \right|~.
\label{eq:shortf}
\ee
The contribution induced by $Q_7^\prime$ is obtained with the replacement 
$C_7 \to \pm C_7^\prime $ in $A^{\rho,\omega}_{\rm PC(PV)}$.

The only $D^0\to V^0\gamma$ decays observed so far are the $K^*$ and $\phi$ modes~\cite{pdg}.
The observed rates satisfy to a good accuracy 
the relation $\cB(D\to K^*\gamma)/\cB(D\to K^*\rho^0) = \cB(D\to \phi\gamma)/ \cB(D\to \phi\rho^0)$,
generally expected by vector meson dominance (VMD). The three 
Cabbibo-suppressed $D^0\to V^0\gamma$ modes, $V^0=\rho^0,\omega,\phi$,
are expected to have similar rates.\footnote{According to explicit VMD predictions~\cite{VMD} $\cB(D\to \rho\gamma)$ and $\cB(D\to \omega\gamma)$
are very similar, possibly a factor $\sim 2$ smaller than $\cB(D^0\to \phi\gamma)$. In the following we 
assume $\cB(D\to (\rho,\omega)\gamma) \geq 10^{-5}$.}
We can thus estimate the typical size of their long-distance amplitudes $|(A^{V}_{\rm PC})^{\rm l.d.}| \simeq |(A^{V}_{\rm PV})^{\rm l.d.}|$ as follows
\bea
\left|(A^{V}_{\rm PC,PV})^{\rm l.d.} \right| & = &
 \left[ \frac{32\pi}{m_D^3} \left(1 - \frac{m_V^2}{m_D^2}\right)^{-3} \frac{\Gamma(D\to V\gamma) }{2} \right]^{1/2}
\nonumber\\  & \to &  \quad \frac{5.8(4)  \times 10^{-8}}{m_D} \quad {\rm for}  \quad V=\phi~.
\label{eq:longf}\
\eea

In the limit where the strong phases of the amplitudes have a mild energy dependence, 
and assuming we can neglect the weak phase of the  long-distance amplitude (see Sec.~\ref{sec:V}),
the direct CP asymmetry, defined in Eq.~(\ref{eq:acpdef}), 
can be decomposed as
\be
|a_{V \gamma}| = 2 ~\zeta_{\rm weak} ~ |\sin(\Delta \phi_{\rm strong})|\,,
\ee
where 
\be 
\zeta_{\rm weak} = \frac{ \left| {\rm Im}(A^{V}_{\rm PC,PV})^{\rm s.d.} \right|} {  \left|  (A^{V}_{\rm PC,PV})^{\rm l.d.} \right| }~.
\ee
As a result, according to Eqs.~(\ref{eq:shortf}) and (\ref{eq:longf}), in the $\rho$ and $\omega$ modes  
the CP violating asymmetries  can reach $10\%$ for maximal strong phases:  
\bea
&& |a_{(\rho,\omega)\gamma}|^{\rm max} = 0.04(1)    \left|\frac{{\rm Im}[C_7(m_c)]} {0.4\times 10^{-2}} \right| \times
\nonumber \\
&& \qquad \times \left[\frac{10^{-5}}{{\mathcal B}(D\to (\rho,\omega)\gamma)}\right]^{1/2}  \lesssim 10\%~. \quad 
\label{eq:maxCPrho} 
\eea

The case of the $\phi$ resonance, or better the  $|K^+K^-\gamma\rangle$ final state with $M_{KK}$ close to the $\phi$ peak, is more 
involved since the hadronic matrix element (\ref{eq:MatQ7}) vanishes,
in the large $m_c$ limit, if $V$ is a pure $s\bar s$ state. However, as we discuss in more detail in 
the next section, a non-negligible CP asymmetry can be expected also in this case 
for two main reasons: 1) the matrix element in (\ref{eq:MatQ7}) is not identically zero even for  $V=\phi$,
both because $O(\Lambda_{QCD}/m_c)$ corrections and because of the tiny $u\bar u$ component of $\phi$;
2) non-resonant contributions due to (off-shell)  $\rho$ and $\omega$ exchange can also
contribute to the $|K^+K^-\gamma\rangle$ final state.

\section{The $D\to K^+K^- \gamma$ case}

The decay amplitudes for $D\to P^+P^-\gamma$ decays can be decomposed in 
full generality as follows 
\bea
&& \cA[ D(p) \to P^+(p_+)P^-(p_-) \gamma(q,\epsilon) ] =  \nonumber \\
&& \quad   -i M(s,\nu)~ \epsilon_{\mu\nu\alpha\beta} q^\mu\epsilon^{*\nu}p^\alpha(p_+-p_-)^\beta   \nonumber \\
&& \quad  +E(s,\nu) ~\epsilon^*_\mu [ q^\mu(qp_+-qp_-)-qp(p_+ -p_-)^\mu ]~, \qquad 
\eea
where $s=(p_++p_-)^2$ and $\nu=(qp_+-qp_-)$. In the limit where we consider at most 
electric and magnetic dipole transitions (or neglecting higher order multipoles), we 
can neglect the $\nu$ dependence of the form factors. In this approximation, 
the differential rate as a function of  $s=M^2_{PP}$ can be written as
\be
\frac{d \Gamma}{ds} = \frac{m_D^3}{32\pi} \left(1 - \frac{s}{m_D^2}\right)^3 \frac{\sqrt{s} \Gamma_0(s)}{\pi} \left[ |M(s)|^2 + |E(s)|^2 \right]~,
\ee
where $\Gamma_0(s)=\sqrt{s}(1-4m_P^2/s)^{3/2}  /(48\pi)$.

If the amplitude is dominated by the exchange of  vector resonances
we can decompose $M$ and $E$ as follows
\bea
M(s) &=& \sum_V \frac{ g^V_{PP} ~A^V_{\rm PC} }{ s - M_V^2 - i \sqrt{s} \Gamma_V  }~, \\
E(s) &=& \sum_V \frac{ g^V_{PP} ~A^V_{\rm PV} }{ s - M_V^2 - i \sqrt{s} \Gamma_V  }~,
\eea
where $g^V_{PP}$ is the $V\to PP$ coupling, defined such that $\Gamma(V\to PP)=g_{PP}^2 \Gamma_0(M_V^2)$.
It is then easy to check that in the limit of a single narrow resonance, integrating over $s$, 
we recover  $\Gamma (D\to PP\gamma) = \Gamma(D\to V\gamma)\times \cB(V\to PP)$.

In order to estimate the maximal direct CP asymmetry in the $D\to K^+K^- \gamma$ case,
with $M_{KK}$ close to the $\phi$ peak, we evaluate $M(s)$ and $E(s)$ summing 
over the three light vector resonances ($V=\rho,\omega,\phi$) with the following assumptions:
\begin{itemize}
\item In all cases we use the parametric form in Eq.~(\ref{eq:longf}) to estimate the overall magnitude 
of $A^V_{\rm PC(PV)}$, assuming further $\cB(D\to (\rho,\omega)\gamma) \geq 10^{-5}$.
\item For $V=\rho,\omega$ we assume the weak phase of $A^V_{\rm PC(PV)}$  is $\zeta_{\rm weak}$, while
for $V=\phi$ we use $r \zeta_{\rm weak}$. Here $r=0.3(1)$ is the typical annihilation suppression factor in non-leptonic $D$ decay 
amplitudes~\cite{Brod:2011re,WA}, that we 
apply to the the matrix element in Eq.~(\ref{eq:MatQ7}) in the  $V=\phi$ case.
\item For $V=\rho,\omega$ we fix the effective coupling to $K^+K^-$  
to $g^V_{K^+K^-}=3$, as expected by $SU(3)$ symmetry given that  $g^\rho_{\pi\pi}\simeq 6$. 
\end{itemize}
Under these hypotheses, and assuming maximal and smoothly varying strong phases for the contributions with different weak phases, we find 
\bea
|a_{K^+K^-\gamma}|^{\rm max}  &\approx&  2\%~, \quad   2m_K <  \sqrt{s} < 1.05~{\rm GeV}~,  \nonumber \\
|a_{K^+K^-\gamma}|^{\rm max}  &\approx&  6\%~,  \quad  1.05~{\rm GeV} <  \sqrt{s} < 1.20~{\rm GeV}~. \nonumber \\
\label{eq:maxCPphi}
\eea
In the first bin, close to the $\phi$ peak, the leading contribution is due to the $\phi$-exchange amplitude.
The contribution due to the non-resonant amplitudes plays a significant role far enough from the $\phi$ peak,
where the charge asymmetry can become larger.  
However, it must be stressed that away from the $\phi$ peak 
the overall rate of the $D\to K^+K^- \gamma$ process is significantly reduced.

\section{Discussion}
\label{sec:V}

In order to establish the significance of these results, two important issues have to be 
clarified:~1) the size of the CP asymmetries within the SM,~2) the role of the strong
phases. 

As far as the SM contribution is concerned, we first notice that short-distance contributions
generated by the operator $Q_7$ are safely negligible: using the result in Ref.~\cite{Greub}
we find  asymmetries below the 0.1\% level. The dominant SM contribution is expected from 
the leading non-leptonic four-quark operators, for which we can apply the general arguments 
presented in~\cite{Isidori:2011qw}. The CP asymmetries can be
decomposed as 
\be
|a^{\rm SM}_{f} |  \approx 2   \xi ~\mathrm{Im} (R_{f}^{\SM}) 
\approx 0.13\% \times \mathrm{Im} (R_{f}^{\SM})~,
\ee
where $\xi \equiv |V_{cb}V_{ub}| / |V_{cs}V_{us}|$
and $R_{f}^{\SM}$ is a ratio of suppressed over leading hadronic amplitudes, 
naturally expected to be smaller than 1. This decomposition holds both 
for the $f=\pi\pi,KK$ channels discussed in Ref.~\cite{Isidori:2011qw}
and for the $f=V\gamma$ case analyzed here. The SM model explanations 
of the result in Eq.~(\ref{eq:acpExp}) require $R_{\pi\pi,KK}^{\SM}\sim3$.
While we cannot exclude this possibility from first principles, 
a further enhancement of one order of magnitude in the $D \to V\gamma$ mode
is beyond any reasonable explanation in QCD. 
As a result, an observation of $|a_{V\gamma}| \gsim 3\%$ would be 
a clear signal of physics beyond the SM, and a clean indication of 
new CP-violating dynamics associated to dipole operators.

Having clarified that large values of $|a_{V\gamma}|$ would 
be a clear footprint of non-standard dipole operators, 
we can ask the question if potential tight limits on $|a_{V\gamma}|$ 
could exclude this non-standard framework. Unfortunately, 
the uncertainty on the strong phases does not allow to draw this conclusion.
We recall that  the maximal values in Eqs.~(\ref{eq:maxCPrho})
and (\ref{eq:maxCPphi}) can be reached only 
in the limit of maximal constructive interference (namely of $\pm \pi/2$ 
strong phase difference) of the amplitudes with different weak phases. 
The calculation of light-quark loop contributions in Ref.~\cite{Greub} does 
suggest the presence of large strong phases in these amplitudes; 
however, we cannot exclude destructive interference effects leading 
to $|a_{V\gamma}| =\mathcal O(0.1\%)$ even in presence of a non-standard CP-violating phase 
in the dipole operator. In principle, this problem could be overcome via 
time-dependent studies of $D(\bar D) \to V\gamma$ decays or using photon polarization, accessible via lepton pair conversion in $D\to V (\gamma^* \to \ell^+\ell^-)$; however,
these types of measurements are certainly more challenging from the 
experimental point of view. \\

\section{Conclusions}

Radiative $D\to P^+P^- \gamma$ decays, with $M_{PP}$ close to the $\rho$ or the $\phi$ peak 
(for $P=\pi$ or $K$, respectively), could help to shed light on the origin of CP violation 
in the charm system. If the experimental  result in Eq.~(\ref{eq:acpExp})
is due to non-standard dynamics involving dipole operators, we can expect   
significantly larger direct CP asymmetries in these radiative modes.
As we have shown, evidence of $|a_{PP\gamma}| \gsim 3\%$ would be 
a clear signal of physics beyond the SM, and a clean indication of 
new CP-violating dynamics associated to dipole operators.

\begin{acknowledgments}

We thank  G.~Giudice and G.~Perez for useful comments and discussions and D. Be\v cirevi\'c for help with hadronic matrix element estimates.
GI acknowledges the support  of the TU~M\"unchen -- Institute for Advanced
Study, funded by the German Excellence Initiative, and the  EU ERC Advanced
Grant FLAVOUR (267104).
The work of JFK was supported in part  by the Slovenian Research Agency. 

\end{acknowledgments}

\end{document}